\documentclass[aps,prd,preprintnumbers,groupedaddress,superscriptaddress,floatfix,tightenlines,twocolumn,reprint,nofootinbib]{revtex4-2}
\usepackage{mathrsfs,natbib}
\usepackage[dvipsnames]{xcolor}
\usepackage{verbatim}
\usepackage{bm,physics}
\usepackage{enumerate}
\usepackage{epsf,amssymb,bbm,amsbsy,amsfonts,amssymb,amsmath}
\usepackage{hyperref}
\usepackage{graphicx}
\usepackage[font=small,labelfont=bf,
   justification=justified,
   format=plain]{caption}

\hypersetup{
    colorlinks=true,       
    linkcolor=red,         
    citecolor=blue,        
    filecolor=magenta,     
    urlcolor=blue          
}

\begin{document}
\title{Criticality of Lower Dimensional AdS\texorpdfstring{$_{d}$}{d} Black Holes}
\author{Aditya Dhumuntarao}
\email{dhumu002@umn.edu}
\affiliation{Perimeter Institute, 31 Caroline St.\ N.\ Waterloo, Ontario, N2L 2Y5, Canada}
\affiliation{School of Physics and Astronomy, University of Minnesota
  Minneapolis, MN 55455, USA}
\author{Robert Mann}
\email{rbmann@uwaterloo.ca}
\affiliation{Perimeter Institute, 31 Caroline St.\ N.\ Waterloo, Ontario, N2L 2Y5, Canada}
\affiliation{Department of Physics and Astronomy, University of Waterloo, Waterloo, Ontario, N2L 3G1, Canada}

\begin{abstract}
	In lower dimensions, charged AdS black holes in an extended phase space, where the cosmological constant is interpreted as the thermodynamic pressure, are typically absent of liquid/gas phase transitions. We investigate the criticality of lower dimensional charged, dilatonic, asymptotically AdS (CDAdS$_d$) black holes generated from consistent truncations of RNAdS$_{d+2}$ black objects. We demonstrate that CDAdS$_{d}$ black holes in $d<4$ can exhibit rich van der Waals behavior and confirm that the associated critical exponents match those expected from mean field theory.
\end{abstract}

\maketitle

\section{Introduction}
	In recent years, black hole chemistry has substantially sharpened the connection between the thermodynamics of AdS black holes and ordinary fluid systems. For rotating, charged black holes with mass $M$, angular momentum $J$, and charge $Q$, it is well known that the first law of black hole mechanics
	\begin{equation}
		\dd M = \frac{\kappa}{8\pi} \dd A + \Omega \dd J + \Phi\dd Q
	\end{equation}
	is identical to the first law of thermodynamics provided one makes the identifications and temperature $T = \kappa/2\pi$, 
 between  surface gravity $\kappa$	and temperature $T$, and~$S = A/4$ between  horizon area $A$ and  entropy $S$ \cite{Bekenstein:1972tm,Bekenstein:1973ur,Hawking:1974rv,Hawking:1974sw}. However, a notable omission in the first law of black hole thermodynamics is a pressure-volume term, which canonically lacks a gravitational interpretation in spacetimes without a cosmological constant $\Lambda$. The fundamental idea of black hole chemistry is to promote the cosmological constant of AdS spacetimes to a thermodynamic variable \cite{Creighton:1995au,Kastor:2009wy,Dolan:2010ha,Dolan:2011jm,Dolan:2011xt}
	\begin{equation}
		P = -\frac{{\Lambda}}{8\pi} 
	\end{equation}
	with a corresponding thermodynamic conjugate volume, $V$ . 
	
	The motivation for this identification is twofold. First, it was observed that the Smarr relation \cite{Smarr:1972kt} is no longer satisfied in spacetimes with a cosmological constant \cite{Kastor:2009wy,Caldarelli:1999xj}. By extending the thermodynamic phase space, the resulting pressure-volume term was found to saturate the Smarr relation. Secondly, the Gibbs free energy of charged Reissner-N\"ordstrom AdS$_d$ (RN-AdS$_d$) black holes has long been known to contain classic swallowtail and cusp behavior suggestive of a van der Waals liquid-gas phase transition \cite{Chamblin:1999tk,Chamblin:1999hg}. The program of black hole chemistry has completed this analogy by demonstrating that the equation of state for RN-AdS$_d$ black holes shares identical critical behavior at the second order transition point and universal compressibility ratio in $d=4$ to a van der Waals fluid \cite{Kubiznak:2012wp,Gunasekaran:2012dq}.
	To date, the correspondence between black hole chemistry and real world fluid systems has yielded an interesting array of thermodynamic features in $d\ge4$, including reentrant phase transitions \cite{Altamirano:2013ane,Gunasekaran:2012dq}, and tricritical points analogous to triple points in the phase diagram of water \cite{Altamirano:2014tva}. 
	
	In contrast to the numerous explorations in $d\ge4$, black hole chemistry in lower dimensions is not as thermodynamically rich. A brief study \cite{Gunasekaran:2012dq} of the charged BTZ$_3$ black hole \cite{Banados:1992gq,Banados:1992wn} and a thorough analysis of lower dimensional black hole chemistry \cite{Frassino:2015oca} revealed that charged black holes in $d=2,3$ have ideal gas like behavior, exhibiting no critical features in the extended phase space. Furthermore, the Smarr relation is explicitly broken for charged BTZ$_3$ black holes unless additional thermodynamic work terms are included in the first law and, in $d=2$, the lack of an area also significantly impacts the canonical definitions of entropy, and thermodynamic volume \cite{Frassino:2015oca}. An outstanding challenge is to find lower dimensional black hole solutions that support non-trivial van der Waals behavior and obey canonical Smarr scalings.

	In this paper, we resolve this tension by identifying thermodynamically rich, charged, dilatonic, asymptotically AdS black hole solutions in $d=2,3$ through dimensional reductions. In Section II, we first review criticality of charged black holes in $d\ge4$ paying close attention to the emergent van der Waals behavior. In Section III, we consider consistent truncations of higher dimensional gravity theories  admitting RNAdS$_{d+2}$ black objects and search for black hole solutions to the effective theory in $d=2,3$. Using lower dimensional techniques in conjunction with Euclidean methods, we explore the thermodynamics of these solutions in the extended phase space. We confirm that the Smarr relation is satisfied in each case and that the equations of state contain local minima indicative of a van der Waals phase transition.
	
\section{PV Criticality of Charged AdS\texorpdfstring{$_d$}{TEXT} Black Holes in \texorpdfstring{$d\ge4$}{TEXT}}

	To set the stage for studying black hole chemistry in lower dimensions, we review some salient features of the Reissener-N\"ordstrom solution to Einstein-Maxwell-AdS$_d$ theory in the extended phase space. We describe the thermodynamics in $d\ge 4$ and study the corresponding equation of state. We then conclude by remarking on the critical behavior and exponents.

	\subsection{Criticality of RN-AdS\texorpdfstring{$_{d}$}{TEXT} Black Holes in \texorpdfstring{$d\ge4$}{TEXT}}
	To demonstrate criticality of charged black holes, the natural starting point is to study the thermodynamics of Einstein-Maxwell-AdS$_{d}$ theory in $d\ge 4$. This theory has been thoroughly analyzed in diverse dimensions with the on-shell regularized Euclidean action first calculated in \cite{Chamblin:1999tk,Chamblin:1999hg}. Explorations of these black holes in an extended phase space were considered in \cite{Kubiznak:2016qmn,Gunasekaran:2012dq}. The bulk action in $d$-dimensions reads 
	\begin{equation}
		\mathcal{I}=-\frac{1}{16\pi G_{d}}\int \dd^{d}x\sqrt{-g}\left[R - 2\Lambda - F^{2}\right] \label{eqn:EMAdS}
	\end{equation}
	with $\Lambda = -(d-2)(d-1)/2\ell^2$ being the cosmological constant. The RN-AdS$_{d}$ solution, in static coordinates, and the purely electric $U(1)$ gauge field are then
	\begin{align}
		\dd s^{2}&=-f(r)\dd t^{2}+\frac{\dd r^{2}}{f(r)}+r^{2}\dd\Omega_{d-2}^{2}  \label{eqn:d+2metric}\\
		F &= \dd A,\hspace{.1in} A = \frac{1}{c}\frac{q}{r^{d-3}}\dd t \label{eqn:gaugefield}
	\end{align}%
	where $\dd \Omega_{n}^{2}$ indicates the metric on the $n$-dimensional sphere, $c=\sqrt{2(d-3)/(d-2)}$, and the metric function has the form
	\begin{equation}\label{eqn:RNAdSd2}
		f(r)=1-\frac{m}{r^{(d-3)}}+\frac{q^{2}}{r^{2(d-3)}}+\frac{r^{2}}{\ell ^{2}}.
	\end{equation}
	Here, the outer horizon $r_+$ is located at the largest positive root of $f(r_+)=0$. The parameters $m$ and $q$ appearing in the metric function are related to the ADM mass and the total charge \cite{Chamblin:1999tk} of the black hole 
	\begin{align}
		M &= \frac{(d-2)\,\omega_{d-2}}{16\pi G_{d}}m,\label{eqn:ADMMass} \\ 
		Q&=\sqrt{2(d-2)(d-3)}\frac{\omega_{d-2}}{8\pi G_{d}}q,\label{eqn:ADMCharge}
	\end{align}
	where $\omega_n$ is the volume of the unit $n$-dimensional sphere,
	\begin{equation}
		\omega_n = \frac{2\pi^{\frac{n+1}{2}}}{\Gamma(\frac{n+1}{2})}.
	\end{equation}
	The standard thermodynamic quantities are then
	\begin{align}
		T&=\frac{f^\prime(r_+)}{4\pi}=\frac{d-3}{4 \pi r_+} +\frac{(d-1) r_+}{4 \pi  \ell ^2}-\frac{(d-3){q}^2 }{4 \pi r_+^{2d-5}}\label{eqn:temp}\\
		S &= \frac{A(\mathcal{H})}{4G_d} = \frac{\omega_{d-2}r^{d-2}_+}{4G_d} \hspace{.8in}\Phi = \frac{1}{c}\frac{q}{r_+^{d-3}}. \label{eqn:RNAdSentropy}
	\end{align}
	In the formalism of the extended phase space \cite{Kubiznak:2016qmn,Kubiznak:2012wp,Gunasekaran:2012dq}, the cosmological constant $\Lambda$ is identified with the thermodynamic pressure of the system via
	\begin{equation}\label{eqn:pressured+2}
		P = -\frac{\Lambda}{8\pi G_d}=\frac{(d-2)(d-1)}{16\pi G_d \ell^2}.
	\end{equation}
	It is then natural to reinterpret the black hole mass as the enthalpy rather than the internal energy and consider a thermodynamic volume  conjugate to the pressure, which reads
	\begin{equation}\label{eqn:vold+2}
		V=\pdv{M}{P}\eval_{S,Q} = \frac{\omega_{d-2}r_+^{d-1}}{d-1}.
	\end{equation}
	In $d>4$, these quantities satisfy the Smarr formula as well as the (extended phase-space) $1st$ law of black hole thermodynamics 
	\begin{align}
		M &= \frac{d-2}{d-3}TS + \Phi\, Q - \frac{2}{d-3}VP\label{eqn:Smarr}\\
		\dd M &= T\,\dd S + \Phi\, \dd Q + V\,\dd P.
	\end{align}
	
	The identification of a pressure term invites a reintrepretation of the black hole temperature, Eqn.~\eqref{eqn:temp}, as an equation of state. This equation of state has been shown to be directly analogous to the equation of state for van der Waals fluids \cite{Kubiznak:2016qmn,Kubiznak:2012wp,Gunasekaran:2012dq}. Comparison of the equation of state, in geometric units $\ell_p^{d-2}=G_d\hbar/c^3$, with a van der Waals fluid in $d$-dimensions reveals that the specific volume $v$ of the fluid and the black hole horizon are related via
	\begin{equation}
		v=\frac{4 r_+ \ell_p^{d-2}}{(d-2)}.
	\end{equation}
	Hence, by employing the definitions of $P$, $T$, and $v$, one finds an equation of state for charged black holes
	\begin{equation}
		P=\frac{T}{v}-\frac{d-3}{(d-2) \pi v^2}+ \frac{(d-3)q^2}{4\pi v^{2(d-2)} \kappa^{2d-5}}\label{eqn:EOS}
	\end{equation}
	where $\kappa = (d-2)/4$. The liquid/gas phase transition experienced by van der Waals fluids is now mapped onto a small-large black hole phase transition in the system for temperatures $T < T_c$ for all $d\ge4.$ The critical points  in the equation of state occur at stationary points of inflection in the $P-v$ diagram given by
	\begin{align}\label{eqn:localmin}
		\frac{\partial P}{\partial v}&=0,& \frac{\partial^2 P}{\partial v^2}&=0.
	\end{align}
	The critical triple $(P_c,T_c,v_c)$ have the values
	\begin{align}
		T_c &=\frac{(d-1)^2}{\pi\kappa v_c(2d-1)},\hspace{.6in}
		P_c =\frac{(d-1)^2}{16\pi \kappa^2 v_c^2},\nonumber\\
		v_c &=\frac{1}{\kappa}\qty[q^2d(2d-1)]^{\frac{1}{2(d-1)}}.\label{eqn:triple}
	\end{align}
	 Analogous to the construction of the universal ratio found for van der Waals fluids, the above triple yields
	\begin{equation}\label{eqn:uniratio}
		\frac{P_c v_c}{T_c}=\frac{2d-5}{4d-8}.
	\end{equation}
	For RN-AdS$_4$, i.e., when $d=4$, one recovers the canonical result of the van der Waals compressibility ratio $P_c v_c/T_c = 3/8$. 

	Given that the equation of state not only reproduces van der Waals behavior but also the universal compressibility ratio, it is natural to consider the critical exponents calculated in the vicinity of the critical point. We begin by considering the entropy, Eqn.~\eqref{eqn:RNAdSentropy}, $S\equiv S(T,V)$, in order to calculate the exponent $\alpha$. Using the volume, Eqn.~\eqref{eqn:vold+2}, the entropy may be written as
	\begin{equation}
		S(T,V) = \omega^{\frac{1}{d-1}}_{d-2}\qty[(d-1)V]^{\frac{(d-2)}{(d-1)}}
	\end{equation}
	where we immediately may deduce that $\alpha= 0$ since the function is independent of $T$. Next, to compute the remaining critical exponents, we define
	\begin{equation}
		p = \frac{P}{P_c},\hspace{.2in}\nu = \frac{v}{v_c}\hspace{.2in}\tau = \frac{T}{T_c}
	\end{equation}
	which translates the equation of state into a `law of corresponding states' analogous to the same function appearing in van der Waals theory, we have
	\begin{equation}\label{eqn:lawofcorrstates}
		p = \frac{4(d-2)}{(2d-5)}\frac{\tau}{\nu} - \frac{(d-2)}{(d-3)\nu^2} + \frac{\nu^{-2(d-2)}}{(d-3)(2d-5)}.	
	\end{equation}
	Note that the only functional dependence on the dimension appears in the last term. As we will demonstrate shortly, the critical exponents are independent of the dimension in $d\ge 4$. This is analogous to the situation in mean field theory where the critical exponents are again independent of the dimension above the upper critical dimension. In this context, this suggests that any equation of state which admits a `law of corresponding states' of the above form will have critical exponents consistent with mean field theory.

	We may consider an expansion of this equation near the critical point 
	\begin{equation}
		\tau = t + 1,\hspace{.3in}\nu = (\omega+1)^{1/z} 	
	\end{equation}
	where $z=d-1>0$. It may be shown that the pressure admits the following expansion 
	\begin{equation}\label{eqn:presseries}
		p = 1 + A_d t - B_dt\omega - C_d\omega^3 +\mathcal{O}(t \omega^2, w^4).
	\end{equation}
	where the coefficients satisfy $A_d,B_d,C_d>0$ and depend on the dimension in the following way
	\begin{equation}
		A_d = \frac{4d-8}{2d-5},\hspace{.08in}B_d = \frac{4d-8}{(2d-5)(d-3)},\hspace{.08in}C_d = \frac{2d-4}{3(d-1)^3}.
	\end{equation}
	Differentiating the series at a fixed $t<0$, we obtain the following differential of the pressure 
	\begin{equation}\label{eqn:pwderivative}
		\dv{p}{\omega} = -(B_d t + 2C_d\omega^2).
	\end{equation}
	Then employing Maxwell's equal area law, see \cite{Kubiznak:2012wp}, the following system of equations emerge for the `volume' of small and large black holes, $\omega_s$ and $\omega_l$ respectively,
	\begin{align}
		p &= 1+ A_d t - B_dt\omega_l - C_d\omega_l^3 = 1 + A_d t - B_dt\omega_s - C_d\omega_s^3\nonumber\\
		0 &= \int_{\omega_l}^{\omega_s} \omega \dd p.
	\end{align}
	The unique solution relates the two volumes via
	\begin{equation}
		\omega_s = -\omega_l = \sqrt{\frac{-Bt}{C}}.
	\end{equation}
	Given these definitions, we are now in a position to derive the critical exponents. First, we have
	\begin{equation}
		\eta = V_c (\omega_l-\omega_s) = 2V_c\omega_l\propto \sqrt{-t} \implies \beta=\frac{1}{2}.
	\end{equation}
	Next, the exponent $\gamma$ can be obtained through the thermodynamic derivative, Eqn.~\eqref{eqn:pwderivative}
	\begin{equation}
		\kappa_T = -\frac{1}{V}\pdv{V}{p}\eval_{T}\propto \frac{1}{Bt} \implies \gamma=1.
	\end{equation}
	Lastly, the `shape of the critical isotherm' is straightforward to obtain by evaluating the series, Eqn.~\eqref{eqn:presseries}, at $t=0$
	\begin{equation}
		p - 1 = -C_d\omega^3 \implies \delta = 3.
	\end{equation}
	Collecting our results, we see that RN-AdS$_d$ black holes have the critical exponents 
	\begin{equation}
		\alpha=0,\hspace{.2in}\beta=\frac{1}{2},\hspace{.2in}\gamma=1,\hspace{.2in}\delta=3
	\end{equation}
	matching precisely with a mean field theory analysis of a van der Waals fluid. This behavior may also be extended to include rotations \cite{Gunasekaran:2012dq}.

	\section{PV Criticality in Lower Dimensions}
	In $d<4$, due to topological restrictions gravity has insufficient symmetry to support PV criticality. To make this observation sharp, consider the equation of state for RN-AdS$_d$ black holes with maximally symmetric horizons in $d\ge 4$
	\begin{align} 
		P=\frac{T}{v} - \frac{(d-3)\hat{k}}{{(d-2)} \pi v^2}+ \frac{(d-3)q^2}{4\pi v^{2(d-2)} \kappa^{2d-5}}\label{eqn:EOSk}
	\end{align}
	where $\hat{k}=+1,0,-1$ for spherical, planar, or hyperbolic horizons respectively. In \cite{Kubiznak:2012wp}, it was first observed that for non-spherically symmetric horizons $(\hat{k}\ne1)$, the equation of state no longer contains a local minimum in which case these black hole solutions do not satisfy the condition Eqn.~\eqref{eqn:localmin} thereby precluding critical behavior. 

	We may compare this equation of state against the spherically symmetric charged BTZ solution \cite{Banados:1992wn,Banados:1992gq} to Einstein-Maxwell-AdS$_{d}$ theory in $d=3$  \cite{Frassino:2015oca}
	\begin{align}
		P &= \frac{T}{v} + \frac{q^2}{2\pi v^2}.\label{eqn:EOSBTZ}
	\end{align}
	We see that, apart from a trivial coefficient, which can be rescaled away, the BTZ equation of state~\eqref{eqn:EOSBTZ}, is analogous to Eqn.~\eqref{eqn:EOSk} with $\hat{k}=0$. In $d=3$, toroidal directions may be freely unwrapped onto the real line by coordinate transformations and hence, the planar $(\hat{k}=0)$ and the spherically symmetric $(\hat{k}=1)$ BTZ solutions share identical equations of state. Clearly, the topological nature of gravity in $d=3$ prevents the charged, BTZ solution to develop local minima. Further analysis of the equation of state of rotating, charged BTZ black holes also confirms a lack of critical behavior \cite{Gunasekaran:2012dq,Frassino:2015oca}. 

	Similar arguments hold for $d=2$ charged black holes solutions to a broad class of dilaton gravity models. In these dimensions, gravity contains no extra spherical directions and the horizon reduces to two points on the real line. It was shown that the equation of state is similar to Eqn.~\eqref{eqn:EOSBTZ} (see Ref.~\cite{Frassino:2015oca} for a detailed review). Clearly, the spherical degrees of freedom associated with an $S^d$, for $d\ge 2$, play a crucial role for non-trivial van der Waals behavior.

	One natural way that lower dimensional gravity can inherit such degrees of freedom is through consistent truncations of higher dimensional gravity theories. Consider the simplest theory of Einstein gravity coupled to a cosmological constant and matter fields in $(d+2)$ dimensions
	\begin{align}\label{eqn:actiond+2}
		\mathcal{I}_{d+2} &= -\frac{1}{16\pi}\int\qty[R_{d+2} + \frac{d(d+1)}{\ell^2}- \mathcal{L}]\boldsymbol{\epsilon}_{d+2}.
	\end{align}
	We assume the above action admits spacetimes that fiber into $\mathcal{M}_d\times S^2$ according to the metric ansatz
	\begin{align}\label{eqn:uplift}
		\dd s^2_{d+2} &= \dd s_{d}^2 + \psi^2\dd\Omega_2^2
	\end{align}
	and that $\psi$ and the matter fields take values along the $d$-dimensional base. By compactifying the spherical directions, one finds the $d$-dimensional truncated theory
	\begin{align}\label{eqn:action2d}
		\mathcal{I} =-\int\qty[\frac{1}{2}(\nabla\psi)^2 + \frac{\psi^2}{4}\qty(R - 2\Lambda - \mathcal{L}) + \frac{\lambda}{2}]\boldsymbol{\epsilon}_d
	\end{align}
	where we have introduced $\Lambda = -d(d+1)/2\ell^2$, and the $d$-dimensional volume form $\boldsymbol{\epsilon}_{d}=\dd^{d}x\sqrt{-g}$. Here, the scalar curvature of the round unit $S^2$ enters the truncated action via $\lambda$ which is set to unity throughout this paper. The spherical directions now descend into lower dimensions via dilaton-gravity couplings. From this reshuffling, we may naively expect that black hole solutions in $d=2,3$ can support non-trivial van der Waals behavior {due to the linear radial divergence of the dilaton field in higher dimensions}. We show that this is the case below.

	The variation of the action with respect to the dilaton and metric yield the equations of motion
	\begin{align}
		0&=\Box \psi - \frac{\psi}{2}\qty(R-2\Lambda-\mathcal{L}),\label{eqn:EOMphi}\\
		0&= \frac{\psi^2}{4}G_{ab} - T_{ab}\label{eqn:EOMmetric}
	\end{align}
	where $G_{ab} = R_{ab}-(R/2) g_{ab}$ is the Einstein tensor and the stress-energy tensor is 
	\begin{align}
		T_{ab} =& -\frac{1}{2}\qty(\nabla_a\psi \nabla_b\psi - \frac{1}{2}g_{ab}[(\nabla\psi)^2-\Lambda \psi^2]) \label{eqn:stressenergy}\\
		&+ \frac{1}{4}\qty[\nabla_a\nabla_b\psi^2-g_{ab}\Box\psi^2] + \frac{\psi^2}{2}T^\mathcal{L}_{ab} + \frac{1}{4} g_{ab}\nonumber.
	\end{align}

	In what follows, we will concentrate on the action Eqn.~\eqref{eqn:action2d} and search for charged, dilatonic, asymptotically AdS$_d$ (CDAdS$_d$) solutions to the equations of motion, Eqns.~\eqref{eqn:EOMphi}--\eqref{eqn:stressenergy}, in $d=2,3$. We compute the thermodynamics using a combination of lower dimensional techniques and the on-shell Euclidean action. Using the extended phase space, we demonstrate that CDAdS$_d$ black hole solutions in $d=2,3$ admit PV criticality and have critical exponents which match those expected from mean field theory.

	\subsection{CDAdS\texorpdfstring{$_2$}{2} BLACK HOLES}
	Here, we consider an Einstein-Maxwell-Dilaton-AdS action with $\mathcal{L}= F^2$ in $d=2$. The uplift of this theory to $4d$ corresponds to Eqn.~\eqref{eqn:EMAdS} for which the RN-AdS$_4$ black hole, Eqn.~\eqref{eqn:d+2metric}, is a solution. We search for the consistently truncated black hole solution in $d=2$ and concentrate on static, purely electric, and radial solutions to the field equations~\eqref{eqn:EOMphi}-\eqref{eqn:stressenergy} with the ansatz
	\begin{align}
		\dd s^2 &= g_{ab}\dd x^a\dd x^b =-f(r)\dd t^2 + f(r)^{-1}\dd r^2,\\
		A&\equiv A(r)\dd t \hspace{.5in} \psi\equiv\psi(r).
	\end{align}
	In $2d$, the Einstein tensor $G_{ab}$ vanishes identically and the Maxwell field strength reduces to $F_{ab} = \epsilon_{ab} \partial_r A(r)$. For the Maxwell Lagrangian, the stress-energy contribution and equation of motion are 
	\begin{align}
		T_{ab}^\mathcal{L} &= F_{ac}F_{b}{}^c - \frac{1}{4}g_{ab}F^2,\hspace{.5in}
		\nabla_a({\psi^2}F^{ab}) = 0.
	\end{align}
	Given this ansatz, the general solution reads
	\begin{align}\label{eqn:CDAdS2}
		f(r) &= 1 - \frac{\Lambda  r^2}{3}-\frac{\mu}{r}+\frac{\rho^2}{r^2},\\
		F &= \dd A, \hspace{.2in} A = -\frac{\rho}{r}\dd t, \hspace{.2in} \psi(r) ={r}.
	\end{align}
	Inserting $\Lambda = -3/\ell^2$, the horizon is given by the largest, real, positive root of 
	\begin{equation}
		f(r_+) = 1 + \frac{r_+^2}{\ell^2} - \frac{\mu}{r_+} + \frac{\rho^2}{r_+^2} = 0.
	\end{equation}
	which imposes $\mu>0.$

	In lower dimensions, computing the thermodynamics requires some care. In $2d$, the action Eqn.~\eqref{eqn:action2d} fits into a broad class of dilaton-gravity models. We follow the procedure outlined in \cite{PhysRevD.47.4438,PhysRevD.52.3494} to determine the mass and charge in terms of $\mu$ and $\rho$. The chemical potential may be computed using standard holographic arguments \cite{Chamblin:1999tk}. The entropy is more subtle, as the area of the horizon is reduced to a point in $2d$. We use the prescription outlined by Wald  which may be formally applied in any dimension $d\ge 2$ \cite{Wald:1993nt,Myers:1994sg}. We compare these quantities against an on-shell evaluation of the Euclidean action to confirm their interpretation and then study PV criticality.

	In $2d$, the current $\mathcal{J}_a\equiv T_{ab}\xi^b$ is conserved regardless of whether the stress-energy is conserved or if $\xi^{a}$ is a Killing vector. Since a divergenceless current is always dual to the gradient of a scalar, it was shown in \cite{PhysRevD.47.4438} that the black hole mass may be computed via
	\begin{equation}
		M = \int \epsilon_{ab} \mathcal{J}^a\dd x^b =  \int \epsilon_{ab} T^{a}{}_{c} \xi^c  \dd x_b.
	\end{equation}
	Using the Killing vector $\xi^a = (1,0)$, and evaluating Eqn.~\eqref{eqn:stressenergy} along Eqn.~\eqref{eqn:CDAdS2}, the mass reads
	\begin{equation}\label{eqn:2dMass}
		M = \frac{\mu}{2}.
	\end{equation}
	
	Next, it has been shown \cite{PhysRevD.52.3494} that for $2d$ dilaton-gravity theories that are coupled to a $U(1)$ gauge field, the total charge may be computed via
	\begin{equation}
		Q = \frac{\psi^2}{\sqrt{-g}} F_{10} = \rho
	\end{equation}
	The corresponding chemical potential is then given by considering the electrostatic potential difference between the horizon and infinity via \cite{Chamblin:1999tk}
	\begin{equation}\label{eqn:2dchempot}
		\Phi = \int_{r_+}^\infty \dd r F_{10} = \frac{\rho}{r_+}.
	\end{equation}
	
	Finally, it is well known that an entropy can be associated with $2d$ black holes by considering an `area' associated with the codimension two surface to the $t-r$ disk \cite{Frassino:2015oca}. A collapsing fluid in $2d$ contains a single point as its boundary on the half line $r>0$; analogously the codimension two boundary of a black hole corresponds to an horizon-point, $r_+$, which has area unity. The Bekenstein-Hawking entropy of the horizon-point is $S_\text{BH} = (1/4)$. However, this entropy conflicts with the one obtained from the Euclidean action.

	An alternative definition of the entropy may be provided by the Wald formula \cite{Wald:1993nt}. In $2d$, the Wald entropy may formally be computed on the codimension two surface to the $t-r$ disk \cite{Myers:1994sg}, i.e., the horizon. We find
	\begin{align}\label{eqn:2dWald}
		S_\text{Wald} = - 2 \pi \fdv{\mathcal{I}}{R^{abcd}} \epsilon_{ac}\epsilon_{bd} \eval_{r_+}= {\pi}r_+^2.
	\end{align}

	In order to confirm the interpretation of these quantities, we consider the on-shell Euclidean action in the canonical ensemble. Rotating the solution into the Euclidean section, $t\to \mathrm{i}\tau$, we find that the counterterm method yields the on-shell Euclidean action (see App.\ref{App:sphericalredux})
	\begin{align}
		\bar{\mathcal{I}} = \frac{\beta}{4}\qty(r_+ - \frac{r^3_+}{\ell^2} + 3\frac{\rho^2}{r_+})
	\end{align}
	where $\beta$ denotes the period the Euclidean section 
	\begin{equation}
		\beta = \frac{4\pi}{\partial_{r}f}\eval_{r=r_+}=\frac{4 \pi  r_+^3 \ell ^2}{(r^2_+-{\rho}^2)\ell^2 + 3r_+^4}.\label{eqn:2dtemp}
	\end{equation}
	The corresponding thermodynamic quantities are
	\begin{align}
		M &= \pdv{\bar{\mathcal{I}}}{\beta}\eval_{{Q}} = \frac{\mu}{2}, \\
		\Phi &= \frac{1}{\beta}\pdv{\bar{\mathcal{I}}}{Q}\eval_{\beta} = \frac{\rho}{r_+},\\  
		S &= \pdv{\bar{\mathcal{I}}}{\beta}\eval_{{Q}} -\bar{\mathcal{I}} = {\pi}{r_+^2}
	\end{align}
	precisely matching Eqns~\eqref{eqn:2dMass}--\eqref{eqn:2dWald}.

	This completes the identification of the standard thermodynamic parameters in $2d$. We now turn our attention to the extended phase space. Asymptotically, $f\to \frac{r^2}{\ell^2}$, and hence the spacetime is AdS$_2$ with the characteristic length scale $\ell$. According to the definitions of the extended phase space, we interpret the pressure as
	\begin{equation}
		P = -\frac{\Lambda}{8\pi} = \frac{3}{8\pi\ell^2}.
	\end{equation}
	Correspondingly,  the conjugate volume may be computed from  
	\begin{equation}
		V = \pdv{M}{P}\eval_{\beta,Q} = \frac{4\pi}{3}r_+^3.
	\end{equation}
	It is then straightforward to check that the first law and the following Smarr relation hold
	\begin{align}
		\dd M &= T \dd S + \Phi \dd Q + V \dd P,\\ 
		M &= 2(T S - V P)+ \Phi Q.
	\end{align}
	The Smarr scaling may be understood by considering the length dimension of mass. From Eqn.~\eqref{eqn:CDAdS2}, the mass has dimension $[M] = L$ due to the quadratic dilaton-gravity couplings in the action. This mass dimension is identical to the RN-AdS$_4$ black hole solution and hence the $d=2$ theory has a similar Smarr scaling.

	The PV equation of state is obtained from expressing the Euclidean period, Eqn.~\eqref{eqn:2dtemp}, in terms of temperature and pressure. In this case, the fluid volume is related to the horizon via $v=2r_+$ and hence the $d=2$ equation of state reads
	\begin{equation}
		P(T,v) =\frac{T}{v} -\frac{1}{2 \pi v^2}+\frac{2{\rho}^2}{ \pi v^4}
	\end{equation}
	which has the critical point
	\begin{equation}
		P_c = \frac{1}{96 \pi  \rho^2}\hspace{.2in} T_c =  \frac{1}{3 \sqrt{6} \pi  \rho} \hspace{.2in} v_c = 2 \sqrt{6} \rho.
	\end{equation}
	The ratio $({P_c v_c}/{T_c}) = ({3}/{8})$ precisely coincides with the universal van der Waals ratio, analogous to the RN-AdS$_4$ case. Hence, we have determined a class of $2d$ charged, black hole solutions to the dilaton-gravity model, Eqn.~\eqref{eqn:action2d}, which supports non-trivial van der Waals behavior. 

	Using the definitions $p = \frac{P}{P_c},$ $\nu = \frac{v}{v_c},$ and $\tau = \frac{T}{T_c}$, we have the `law of corresponding states' 
	\begin{equation}
		p(\tau,\nu) = \frac{8}{3}\frac{\tau}{\nu} - \frac{2}{\nu^2} + \frac{1}{3\nu^{4}}.
	\end{equation}
	Following the analysis given above, it is straightforward to conclude that the critical exponents are those of mean field theory;
	\begin{equation}
		\alpha=0,\hspace{.2in}\beta=\frac{1}{2},\hspace{.2in}\gamma=1,\hspace{.2in}\delta=3.
	\end{equation}	
	
	\subsection{{CDAdS\texorpdfstring{$_3$}{3} BLACK HOLES}}

	We now turn our attention to $3d$ black hole solutions that result from consistent truncations of the bulk action in $5d$, Eqn.~\eqref{eqn:actiond+2}.  It was shown in \cite{Cisterna:2019scr} that homogeneous uncharged, AdS$_d$ black strings are permissible geometries that solve the field equations of the action~\eqref{eqn:actiond+2} with $\mathcal{L} = (\nabla \Sigma)^2$. The massless scalar field modifies the cosmological constant to support the black string along flat directions. Here, we construct charged, dilatonic, AdS$_3$ planar black holes using $\mathcal{L} = F^2 + (\nabla \Sigma)^2$. The uplift of this solution to $d=5$ dimensions corresponds to charged AdS$_5$ black strings with horizon topologies $S^2\times \mathbb{R}^1$.

	We search for planar $3d$ black hole solutions with purely electric solutions to the field equations~\eqref{eqn:EOMphi}--\eqref{eqn:stressenergy} with the ansatz 
	\begin{align}
		\dd s^2_3 &= -f(r)\dd t^2 + f(r)^{-1}\dd r^2 + h(r)^2\dd x^2\label{eqn:metric3d}\\
		A &\equiv A(r)\dd t\hspace{.25in} \psi\equiv\psi(r) \hspace{.25in} \Sigma\equiv\Sigma(x)
	\end{align}
	In this case, the equations of motion for the gauge and scalar fields are
	\begin{equation}
		\nabla_a(\psi^2 F^{ab}) = 0, \hspace{.5in} \nabla_a(\psi^2\nabla^a\Sigma)=0
	\end{equation}
	and the corresponding stress-energy contribution is
	\begin{equation}
		T_{ab}^\mathcal{L} = F_{ac}F_b{}^c - \frac{1}{4}g_{ab} F^2 - \nabla_a\Sigma\nabla_b\Sigma + \frac{1}{2}g_{ab}(\nabla\Sigma)^2.\label{eqn:matter3dstress}
	\end{equation}
	Given this ansatz, it is straightforward to demonstrate that the equations of motion are satisfied for the solution
	\begin{align}\label{eqn:3DSCBH}
		f(r) &=\frac{1}{2\varphi^2}-\frac{\Lambda  r^2}{6}-\frac{\mu }{r^2}+\frac{\rho^2}{r^4}, \\
		\psi(r) &= \varphi r, \hspace{.8in} h(r) = r,\\
		A &= -\sqrt{\frac{3}{4}}\frac{\rho}{r^2} \dd t,\hspace{.35in} {\Sigma(x)} = \frac{x}{\sqrt{2}\varphi}.
	\end{align}
	Here, the horizon is given by the largest positive root of 
	\begin{equation}
		f(r_+) = \frac{1}{2\varphi^2} + \frac{r_+^2}{\ell^2} - \frac{\mu}{r_+^2} + \frac{\rho^2}{r_+^4} = 0
	\end{equation}
	where we have used $\Lambda = -6/\ell^2$. Note that for a real root, the bound $\mu>0$ must be saturated. The free parameter is set by demanding $2\varphi^2=1$.

	As we will demonstrate, the parameters $\mu$ and $\rho$ are again respectively related to the black hole mass and charge. When computing the thermodynamics, the presence of a planar horizon leads to infinities. To circumvent this, we consider thermodynamic densities with respect to the planar direction.

	We first compute the ADM mass using the the background subtraction procedure outlined in \cite{Chan:1996sx,Hawking:1995fd}. The total charge is given by the weighted flux across the planar horizon and the chemical potential is determined by considering the electrostatic potential difference from the horizon and infinity \cite{Chamblin:1999tk}. In $3d$, the spherically symmetric horizon allows for a direct computation of both the Bekenstein-Hawking and Wald entropies. We check these quantities against the on-shell Euclidean action and identify the correct entropy before studying the PV behavior.

	To determine the ADM mass, we use the background subtraction procedure outlined in \cite{Chan:1996sx,Hawking:1995fd}. In $3d$, the mass function reads
	\begin{align}
		\mathcal{M}(r)&= {2\sqrt{f}}\qty(\sqrt{f_\text{ref}}-\sqrt{f})\dv{r} \qty(\frac{\psi^2}{4}\mathcal{A})
	\end{align}
	where $\mathcal{A} = x\, r$ is the area function of planar horizon. In this case, the reference spacetime 
	\begin{equation}
		\dd s_\text{ref}^2 = -\qty(1 + \frac{r^2}{\ell^2})\dd t^2 + \frac{\dd r^2}{\qty(1 + \frac{r^2}{\ell^2})} + r^2 \dd x^2
	\end{equation}
	is simply the asymptotically\footnote{Provided the toroidal identification $x\sim x+2\pi$, the space is locally AdS$_3$ at the expense of a discontinuity in $\Sigma(x)$.} AdS$_3$ solution obtained by $\mu=\rho=0$. The ADM mass density, computed at the radial boundary, is
	\begin{align}
		m &= \lim_{r\to\infty}\frac{1}{x} \mathcal{M}(r)= \frac{3\mu }{8}.\label{eqn:3dMass}
	\end{align}

	For our purposes, it is important to compute the total charge using a weighted flux across the planar horizon. To motivate this calculation, we consider the charge of the uplifted solution. In $5d$, the electric flux across the $S^2\times\mathbb{R}$ horizon determines the total charge of an RN-AdS$_5$ black string. Inserting Eqn.~\eqref{eqn:metric3d} into Eqn.~\eqref{eqn:uplift}, the total charge reads
	\begin{equation}
		Q = \frac{1}{4\pi}\int \star F = \frac{1}{4\pi}\int F_{10} (\psi^2 h)\dd\Omega\dd x.
	\end{equation}
	Integrating out the spherical directions, we interpret the charge density in $3d$ as
	\begin{equation}
		q = \frac{1}{x} \int \star (\psi^2 F) =  \frac{\sqrt{3}}{2}\rho.\label{eqn:3dCharge}
	\end{equation}
	The corresponding chemical potential follows the standard definition \cite{Chamblin:1999tk},
	\begin{equation}
		\Phi = \int_{r_+}^\infty \dd r F_{10} = \sqrt{\frac{3}{4}}\frac{\rho}{r_+^2}.\label{eqn:3dchempot}
	\end{equation}

	In $3d$, we have the freedom to either consider the Bekenstein-Hawking entropy or the Wald entropy. Using the standard formulas, one has the entropy densities
	\begin{align}
		s_\text{BH} &= \frac{1}{x}\frac{\mathcal{A}}{4} = \frac{r_+}{4},\\
		s_\text{Wald} &= \frac{2\pi}{x} \int \qty(\fdv{\mathcal{I}}{R^{abcd}} \epsilon^{ac}\epsilon^{bd}) \dd x = \frac{\pi r_+^3}{2}.\label{eqn:3dWald}
	\end{align}
	To single out the correct thermal entropy, once again we can appeal to the on-shell Euclidean action. Using the counterterm subtraction method \cite{Emparan:1999pm}, the on-shell Euclidean action in the canonical ensemble reads
	\begin{equation}
		\bar{\mathcal{I}} =\frac{\beta x}{8\ell^2} \qty( \frac{r_+^4 \ell ^2-r_+^6+5 \rho^2 \ell ^2}{r_+^2})
	\end{equation}
	where $\beta$ denotes the period of the Euclidean section
	\begin{equation}\label{eqn:3dtemp}
		\beta = \frac{4\pi}{\partial_r f}\eval_{r=r_+} = \frac{2 \pi  r_+^5 \ell ^2}{(r_+^4-\rho^2) \ell ^2+2 r_+^6}.
	\end{equation}
	In the fixed charge ensemble, the mass, chemical potential, and entropy are straightforward to compute
	\begin{align}
		m &= \pdv{\bar{i}}{\beta}\eval_{Q} = \frac{3\mu}{8}, \\
		\Phi &= \frac{1}{\beta} \pdv{\bar{i}}{Q}\eval_{\beta} = \sqrt{\frac{3}{4}} \frac{\rho}{r_+^2},\\ 
		s &= \pdv{\bar{i}}{\beta}\eval_{Q} -\bar{i} = \frac{\pi r_+^3}{2\pi}, 
	\end{align}
	where we have used the action density $\bar{i} = (\bar{\mathcal{I}}/x)$. Comparing these quantities against Eqn.~\eqref{eqn:3dMass}, Eqn.~\eqref{eqn:3dchempot}, and Eqn.~\eqref{eqn:3dWald} respectively confirms the interpretation of the charge and selects the Wald entropy as the correct thermodynamic entropy.

	We are now in the position to study PV criticality. Clearly, the spacetime is asymptotically AdS$_3$ with respect to the length scale $\ell$, and hence the pressure in the extended phase space formalism may be identified with
	\begin{align}
		P &= - \frac{\Lambda}{8\pi} = \frac{3}{4\pi \ell^2}.
	\end{align}
	In the fixed charge ensemble, this invites a reinterpretation of the ADM mass as the enthalpy $m\equiv m(P,S)$ and we can compute the thermodynamic volume 
	\begin{align}
		V &= \pdv{m}{P}\eval_{S,Q} = {\pi^2r_+^4}.
	\end{align}
One can check that the first law and  Smarr relation 
	\begin{align}
		\dd M &= T \dd S + \Phi \dd Q + V \dd P \\
		M &= \frac{3}{2}TS - VP + \Phi Q. 
	\end{align}
hold 	with these identifications.
	The corresponding equation of state may be found from inverting the periodicity, Eqn.~\eqref{eqn:3dtemp}, and using the definition of the fluid volume. In this case, the fluid volume is related to the horizon via $v = 4r_+\ell_p/3$ resulting in the equation of state
	\begin{equation}
		P(T,v) = \frac{T}{v}-\frac{2}{3 \pi  v^2}+\frac{512 \rho ^2}{243 \pi  v^6}
	\end{equation}
	where we have set $\ell_p=1.$ The critical point of the equation of state yields the triple
	\begin{equation}
		P_c= \frac{1}{4 \sqrt{15} \pi  \rho }, \hspace{.2in} T_c= \frac{4}{5\pi ({15\rho^2})^{\frac{1}{4}}},\hspace{.2in} v_c ={4 \qty({{5\rho^2}/{3} })^{\frac{1}{4}} }.
	\end{equation}
	and satisfies the ratio ${P_c v_c}/{T_c} = {5}/{12}$. Note that this ratio is also shared by RN-AdS$_5$ black holes, Eqn.~\eqref{eqn:uniratio}. This suggests that the charged AdS$_5$ black string uplift of CDAdS$_3$ black holes belongs in the same universality class as RN-AdS$_5$ black holes.

	Constructing the law of corresponding states follows the same procedure as in higher dimensions. Employing the definitions $p = \frac{P}{P_c},$ $\nu = \frac{v}{v_c},$ and $\tau = \frac{T}{T_c}$, the law of corresponding states can be written as
	\begin{equation}
		p = \frac{12\tau}{5\nu} - \frac{3}{2\nu^2} + \frac{1}{10v^{6}}
	\end{equation}
	which coincides precisely Eqn.~\eqref{eqn:lawofcorrstates} for $d=5$. Hence, we conclude that the critical exponents are again consistent with mean field theory and this completes the identification of lower dimensional black holes demonstrating critical behavior.

	\section{Conclusions}
	
	Given the lack of lower dimensional charged black hole solutions  admitting van der Waals behavior, we constructed gravity theories in $d=2,3$ via consistent truncations and explored the extended phase space of black hole solutions using lower dimensional techniques and Euclidean methods. We have found classic van der Waals structures in the equation of state of charged, dilatonic, asymptotically AdS black hole solutions in $d=2,3$ and confirmed that the critical exponents match those expected from mean field theory.

	In $d=2$, the mass and charge of the CDAdS$_2$ solution, Eqn.~\eqref{eqn:CDAdS2}, were straightforward to compute with lower dimensional methods outlined in \cite{PhysRevD.47.4438,PhysRevD.52.3494}. We found, rather than the standard definition of Bekenstein-Hawking entropy, the Wald entropy yields an expression that agrees with the entropy calculated from the on-shell Euclidean action. Furthermore, this solution has several features reminiscent of RN-AdS$_4$ black hole solutions of Einstein-Maxwell-AdS$_4$ theory, especially the similarity between the respective equations of state and the coincident van der Waals ratios. The CDAdS$_2$ solution, however, does not satisfy the canonical Smarr relation in $d=2$ but instead saturates the formula with $d=4$. This is due to the quadratic dilaton-gravity coupling in the action, which raises the mass dimension of the black hole to that of $4d$.

	In $d=3$, searching for black hole solutions required some care. We performed consistent truncations of theories  admitting charged AdS$_5$ black strings with horizon topologies $S^2\times \mathbb{R}$ along the spherical directions. The dimensionally reduced theory had an additional scalar field that supported planar charged, asymptotically AdS black strings in $d=3$. To regulate infinites arising from the planar direction, we computed the thermodynamic densities and confirmed that the mass and charge density have a standard interpretation. Again, the Wald entropy density matches   the entropy density computed from the on-shell Euclidean action, confirming that the Wald entropy, rather than the Bekenstein-Hawking entropy, is the correct thermal entropy. Similar to the case in $d=2$, the canonical Smarr relation was not satisfied with $d=3$. Instead, the scaling properties of these black hole solutions were identical to RN-AdS$_5$ black strings. The Smarr scaling in our work contrasts an earlier analysis \cite{Frassino:2015oca} where new work terms were required to satisfy Smarr laws for Einstein-Maxwell-AdS$_3$ theory without additional scalar fields. We confirmed that the equation of state contains critical phenomena and the van der Waals ratio is identical to that of RN-AdS$_5$ black holes. The latter observation suggests that compactification of black objects with various horizon topologies preserves the van der Waals universality class.
	
	There are several interesting directions to pursue. We were here primarily concerned with searching for rich van der Waals behavior in lower dimensions with a dynamical scalar field. However other phase transitions may be available. Indeed by imposing a constant scalar field, the action reduces to 
	\begin{equation}
		\mathcal{I}[\psi_0] = - \int\qty[\frac{\psi_0^2}{4}(R-2\Lambda-\mathcal{L})+\frac{\lambda}{2} ] \boldsymbol{\epsilon}_d
	\end{equation}
	with an effective cosmological constant $\Lambda_\text{eff} = \frac{\psi_0^2}{4}\Lambda + \frac{-\lambda}{4}.$ The action in $d=3$ has known black hole solutions, namely the charged BTZ solution with a planar horizon. Hence  the scalar behaves analogous to an order parameter that supports the CDAdS$_3$ string phase with pressure corresponding to $\Lambda$,  or  to the charged BTZ$_3$ string phase with pressure given by $\Lambda_\text{eff}$. It would be fruitful to explore an ensemble which allows for such Landau-Ginzburg phase transitions and study if the CDAdS$_3$ solution lies within the BTZ mass gap. 
	
	Similar arguments can be made for the action in $d=2$. In this case, it is convenient to recall that the near horizon extremal geometry of an RN-AdS$_4$ black hole fibers into AdS$_2\times S^2_{\psi_0}$, i.e, the Robinson-Bertolli spacetime. Upon compactification, this spacetime should appear as a classical solution to the above action if one treats $\psi_0$ as a Lagrange multiplier. Again the dynamics of the scalar provides access to different black hole phases of the theory. It would be very interesting to precisely chart out the complete phase space of solutions to the general compactified action~\eqref{eqn:action2d}. 
	
	Finally, another possible direction is to consider our approach in $d>3$. Here several new features are available, such as magnetic black hole solutions, $p$-form electromagnetism, Gauss-Bonnet and Lovelock sectors of gravity, and novel phase transitions in addition to van der Waals. 

	\section*{Acknowledgments}
	This work was supported in part by the Natural Sciences and Engineering Research Council of Canada. AD is further supported by the National Science Foundation Graduate Research Fellowship under Grant No.\ 00039202. This research was supported in part by the Perimeter Institute for Theoretical Physics. Research at Perimeter Institute is supported by the Government of Canada through the Department of Innovation, Science and Economic Development Canada and by the Province of Ontario through the Ministry of Research, Innovation and Science.
 	
	\newpage
	\appendix
	\section{Spherical Reduction}\label{App:sphericalredux}
	
	We review the dimensional reduction procedure we employ  in detail. Consider Einstein gravity coupled to a cosmological constant with additional matter degrees of freedom. The total action is
	\begin{align}
		\mathcal{I} &= -\frac{1}{8\pi}\times\qty(\mathcal{I}_{d+2}+\mathcal{I}_\text{surf}-\mathcal{I}_\text{ct})
	\end{align}
	with $G_{d+2}$ is the gravitational constant in $(d+2)$ dimensions and the bulk action is given by
	\begin{align}
		\mathcal{I}_{d+2}&=\frac{1}{2}\int \qty[R - 2\Lambda - \mathcal{L}]\boldsymbol{\epsilon}_{d+2}
	\end{align}
	where $\mathcal{L}$ is the matter Lagrangian, $\Lambda$ is the cosmological constant in $(d+2)$-dimensions, and $\boldsymbol{\epsilon}_{d+2}$ is the volume form on the spacetime. The remaining surface and counterterm actions (up to $d=6$) are defined on the spacelike codimension one boundary,
	\begin{align}
		\mathcal{I}_\text{surf}=& \int\qty[\mathcal{K}+ 2n_a F^{ab}A_b]\boldsymbol{\epsilon}_{d+1}\\
		\mathcal{I}_\text{ct}=&\int\left[\frac{d}{\ell} + \frac{\ell c_1(d)}{2(d-1)}\mathcal{R} \right.\\
		&\left.+\frac{\ell^3c_2(d)}{2(d-1)^2}\qty(\mathcal{R}^{ab}\mathcal{R}_{ab}- \frac{d+1}{4d}\mathcal{R}^2)+\cdots\right]\boldsymbol{\epsilon}_{d+1}\nonumber
	\end{align}
	which are required both for a well-posed variational principle, and to regularize the Euclidean action in the canonical ensemble \cite{Emparan:1999pm,Chamblin:1999tk,Kubiznak:2012wp}. The coefficients $c_i(d)$ are present to regulate Casimir contributions when computing the on-shell action for black string geometries.

	We consider metrics of the form
	\begin{align}
		\dd s^2_{d+2} &=  g_{ab}\dd x^a\dd x^b+ \psi(x^a)^2\dd\Omega_2^2.
	\end{align}
	For $d=2$ and $\mathcal{L}=F^2$, the static, spherically symmetric, charged black hole solution is the RN-AdS$_4$ solution with $\psi = r.$ Note that when $d=3$, the horizon has the topology $S^2\times S^1$ or $S^2\times \mathbb{R}$ corresponding to a black string. Recently, homogeneous asymptotically AdS$_{n}\times \mathbb{R}^p$ black strings have been stabilized using $p$ minimally coupled scalar fields that depend on the extended $p$-directions \cite{Cisterna:2019scr}.  We consider here $p=1$ in order to support inhomogeneous, asymptotically AdS, charged black strings in $d=3$ which admit horizon topologies $S^2\times \mathbb{R}$ (or $S^1$).

	We are now in a position to perform a consistent truncation of the theory along the $S^2$ of the horizon to $d$-dimensions. To do so, one further requires the data
	\begin{align}
		\boldsymbol{\epsilon}_{d+2} &= \psi^2\det(\Omega_2)\boldsymbol{\epsilon}_d,\\
		R_{d+2} &= R_{d} + \frac{2}{\psi^2} - \frac{4\square\psi}{\psi} - \frac{2(\nabla\psi)^2}{\psi^2}, \\
		\boldsymbol{\epsilon}_{d+1} &= \psi^2\det(\Omega_2)\boldsymbol{\epsilon}_{d-1}, \\
		\mathcal{K}_{d+2} &= \mathcal{K}_{d} + \frac{2}{\psi}\sqrt{-g}\, {n^a \nabla_a\psi},\\
		\mathcal{R}^{ab}\mathcal{R}_{ab} &= \frac{1}{2} \mathcal{R}^2 = \frac{2}{\psi^4}.
	\end{align}
	where the derivative operators are computed on the base $g_{ab}$. Using this data, the spherical directions and the surface term in $R_{d+2}$ can be integrated away. The effective bulk theory is 
	\begin{align}
		\mathcal{I} &=-\int\qty[\frac{1}{2}(\nabla\psi)^2 + \frac{\psi^2}{4}\qty(R - 2\Lambda - \mathcal{L}) +\frac{\lambda}{2} ]\boldsymbol{\epsilon}_d.
	\end{align}
	where $\lambda = 1$ is related to the curvature of the round unit $S^2$. The remaining surface and counterterm actions 
	\begin{align}
		\mathcal{I}_\text{surf} = -\frac{1}{2G_d}&\int {\psi^2}\qty[\mathcal{K}_h + 2n_a F^{ab}A_b]\boldsymbol{\epsilon}_{d-1},\\
		\mathcal{I}_\text{ct} = +\frac{1}{2G_d}&\int{\psi^2}\left[\frac{d}{\ell}+ \frac{\ell}{2(d-1)}\frac{c_1(d)}{\psi^2}\right.\nonumber \\
		& + \frac{\ell ^3}{2 d({d}-1)^2}\frac{c_2(d)}{\psi^4}]\boldsymbol{\epsilon}_{d-1}.
	\end{align}
	guarantee that the total action is finite when computing the value on-shell. For $d=2,3$, the coefficients take the form
	\begin{equation}
		c_1(d) = \frac{d}{2},\hspace{.3in} c_2(d) = -\qty(\frac{d-2}{2})\qty(\frac{d}{d-1})^2.
	\end{equation}
	This choice naturally eliminates any additional Casimir contributions along the non-compact directions for $d\le6$.


\bibliographystyle{apsrev4-1}
\bibliography{DMReferences}

\begin{thebibliography}{29}%
\makeatletter
\providecommand \@ifxundefined [1]{%
 \@ifx{#1\undefined}
}%
\providecommand \@ifnum [1]{%
 \ifnum #1\expandafter \@firstoftwo
 \else \expandafter \@secondoftwo
 \fi
}%
\providecommand \@ifx [1]{%
 \ifx #1\expandafter \@firstoftwo
 \else \expandafter \@secondoftwo
 \fi
}%
\providecommand \natexlab [1]{#1}%
\providecommand \enquote  [1]{``#1''}%
\providecommand \bibnamefont  [1]{#1}%
\providecommand \bibfnamefont [1]{#1}%
\providecommand \citenamefont [1]{#1}%
\providecommand \href@noop [0]{\@secondoftwo}%
\providecommand \href [0]{\begingroup \@sanitize@url \@href}%
\providecommand \@href[1]{\@@startlink{#1}\@@href}%
\providecommand \@@href[1]{\endgroup#1\@@endlink}%
\providecommand \@sanitize@url [0]{\catcode `\\12\catcode `\$12\catcode
  `\&12\catcode `\#12\catcode `\^12\catcode `\_12\catcode `\%12\relax}%
\providecommand \@@startlink[1]{}%
\providecommand \@@endlink[0]{}%
\providecommand \url  [0]{\begingroup\@sanitize@url \@url }%
\providecommand \@url [1]{\endgroup\@href {#1}{\urlprefix }}%
\providecommand \urlprefix  [0]{URL }%
\providecommand \Eprint [0]{\href }%
\providecommand \doibase [0]{http://dx.doi.org/}%
\providecommand \selectlanguage [0]{\@gobble}%
\providecommand \bibinfo  [0]{\@secondoftwo}%
\providecommand \bibfield  [0]{\@secondoftwo}%
\providecommand \translation [1]{[#1]}%
\providecommand \BibitemOpen [0]{}%
\providecommand \bibitemStop [0]{}%
\providecommand \bibitemNoStop [0]{.\EOS\space}%
\providecommand \EOS [0]{\spacefactor3000\relax}%
\providecommand \BibitemShut  [1]{\csname bibitem#1\endcsname}%
\let\auto@bib@innerbib\@empty
\bibitem [{\citenamefont {Bekenstein}(1972)}]{Bekenstein:1972tm}%
  \BibitemOpen
  \bibfield  {author} {\bibinfo {author} {\bibfnamefont {J.~D.}\ \bibnamefont
  {Bekenstein}},\ }\href {\doibase 10.1007/BF02757029} {\bibfield  {journal}
  {\bibinfo  {journal} {Lett. Nuovo Cim.}\ }\textbf {\bibinfo {volume} {4}},\
  \bibinfo {pages} {737} (\bibinfo {year} {1972})}\BibitemShut {NoStop}%
\bibitem [{\citenamefont {Bekenstein}(1973)}]{Bekenstein:1973ur}%
  \BibitemOpen
  \bibfield  {author} {\bibinfo {author} {\bibfnamefont {J.~D.}\ \bibnamefont
  {Bekenstein}},\ }\href {\doibase 10.1103/PhysRevD.7.2333} {\bibfield
  {journal} {\bibinfo  {journal} {Phys. Rev. D}\ }\textbf {\bibinfo {volume}
  {7}},\ \bibinfo {pages} {2333} (\bibinfo {year} {1973})}\BibitemShut
  {NoStop}%
\bibitem [{\citenamefont {Hawking}(1974)}]{Hawking:1974rv}%
  \BibitemOpen
  \bibfield  {author} {\bibinfo {author} {\bibfnamefont {S.~W.}\ \bibnamefont
  {Hawking}},\ }\href {\doibase 10.1038/248030a0} {\bibfield  {journal}
  {\bibinfo  {journal} {Nature}\ }\textbf {\bibinfo {volume} {248}},\ \bibinfo
  {pages} {30} (\bibinfo {year} {1974})}\BibitemShut {NoStop}%
\bibitem [{\citenamefont {Hawking}(1975)}]{Hawking:1974sw}%
  \BibitemOpen
  \bibfield  {author} {\bibinfo {author} {\bibfnamefont {S.~W.}\ \bibnamefont
  {Hawking}},\ }\href {\doibase 10.1007/BF02345020} {\bibfield  {journal}
  {\bibinfo  {journal} {Commun. Math. Phys.}\ }\textbf {\bibinfo {volume}
  {43}},\ \bibinfo {pages} {199} (\bibinfo {year} {1975})},\ \bibinfo {note}
  {[Erratum: Commun.Math.Phys. 46, 206 (1976)]}\BibitemShut {NoStop}%
\bibitem [{\citenamefont {Creighton}\ and\ \citenamefont
  {Mann}(1995)}]{Creighton:1995au}%
  \BibitemOpen
  \bibfield  {author} {\bibinfo {author} {\bibfnamefont {J.~D.~E.}\
  \bibnamefont {Creighton}}\ and\ \bibinfo {author} {\bibfnamefont {R.~B.}\
  \bibnamefont {Mann}},\ }\href {\doibase 10.1103/PhysRevD.52.4569} {\bibfield
  {journal} {\bibinfo  {journal} {Phys. Rev. D}\ }\textbf {\bibinfo {volume}
  {52}},\ \bibinfo {pages} {4569} (\bibinfo {year} {1995})},\ \Eprint
  {http://arxiv.org/abs/gr-qc/9505007} {arXiv:gr-qc/9505007} \BibitemShut
  {NoStop}%
\bibitem [{\citenamefont {Kastor}\ \emph {et~al.}(2009)\citenamefont {Kastor},
  \citenamefont {Ray},\ and\ \citenamefont {Traschen}}]{Kastor:2009wy}%
  \BibitemOpen
  \bibfield  {author} {\bibinfo {author} {\bibfnamefont {D.}~\bibnamefont
  {Kastor}}, \bibinfo {author} {\bibfnamefont {S.}~\bibnamefont {Ray}}, \ and\
  \bibinfo {author} {\bibfnamefont {J.}~\bibnamefont {Traschen}},\ }\href
  {\doibase 10.1088/0264-9381/26/19/195011} {\bibfield  {journal} {\bibinfo
  {journal} {Class. Quant. Grav.}\ }\textbf {\bibinfo {volume} {26}},\ \bibinfo
  {pages} {195011} (\bibinfo {year} {2009})},\ \Eprint
  {http://arxiv.org/abs/0904.2765} {arXiv:0904.2765 [hep-th]} \BibitemShut
  {NoStop}%
\bibitem [{\citenamefont {Dolan}(2011{\natexlab{a}})}]{Dolan:2010ha}%
  \BibitemOpen
  \bibfield  {author} {\bibinfo {author} {\bibfnamefont {B.~P.}\ \bibnamefont
  {Dolan}},\ }\href {\doibase 10.1088/0264-9381/28/12/125020} {\bibfield
  {journal} {\bibinfo  {journal} {Class. Quant. Grav.}\ }\textbf {\bibinfo
  {volume} {28}},\ \bibinfo {pages} {125020} (\bibinfo {year}
  {2011}{\natexlab{a}})},\ \Eprint {http://arxiv.org/abs/1008.5023}
  {arXiv:1008.5023 [gr-qc]} \BibitemShut {NoStop}%
\bibitem [{\citenamefont {Dolan}(2011{\natexlab{b}})}]{Dolan:2011jm}%
  \BibitemOpen
  \bibfield  {author} {\bibinfo {author} {\bibfnamefont {B.~P.}\ \bibnamefont
  {Dolan}},\ }\href {\doibase 10.1103/PhysRevD.84.127503} {\bibfield  {journal}
  {\bibinfo  {journal} {Phys. Rev.}\ }\textbf {\bibinfo {volume} {D84}},\
  \bibinfo {pages} {127503} (\bibinfo {year} {2011}{\natexlab{b}})},\ \Eprint
  {http://arxiv.org/abs/1109.0198} {arXiv:1109.0198 [gr-qc]} \BibitemShut
  {NoStop}%
\bibitem [{\citenamefont {Dolan}(2011{\natexlab{c}})}]{Dolan:2011xt}%
  \BibitemOpen
  \bibfield  {author} {\bibinfo {author} {\bibfnamefont {B.~P.}\ \bibnamefont
  {Dolan}},\ }\href {\doibase 10.1088/0264-9381/28/23/235017} {\bibfield
  {journal} {\bibinfo  {journal} {Class. Quant. Grav.}\ }\textbf {\bibinfo
  {volume} {28}},\ \bibinfo {pages} {235017} (\bibinfo {year}
  {2011}{\natexlab{c}})},\ \Eprint {http://arxiv.org/abs/1106.6260}
  {arXiv:1106.6260 [gr-qc]} \BibitemShut {NoStop}%
\bibitem [{\citenamefont {Smarr}(1973)}]{Smarr:1972kt}%
  \BibitemOpen
  \bibfield  {author} {\bibinfo {author} {\bibfnamefont {L.}~\bibnamefont
  {Smarr}},\ }\href {\doibase 10.1103/PhysRevLett.30.521,
  10.1103/PhysRevLett.30.71} {\bibfield  {journal} {\bibinfo  {journal} {Phys.
  Rev. Lett.}\ }\textbf {\bibinfo {volume} {30}},\ \bibinfo {pages} {71}
  (\bibinfo {year} {1973})},\ \bibinfo {note} {[Erratum: Phys. Rev.
  Lett.30,521(1973)]}\BibitemShut {NoStop}%
\bibitem [{\citenamefont {Caldarelli}\ \emph {et~al.}(2000)\citenamefont
  {Caldarelli}, \citenamefont {Cognola},\ and\ \citenamefont
  {Klemm}}]{Caldarelli:1999xj}%
  \BibitemOpen
  \bibfield  {author} {\bibinfo {author} {\bibfnamefont {M.~M.}\ \bibnamefont
  {Caldarelli}}, \bibinfo {author} {\bibfnamefont {G.}~\bibnamefont {Cognola}},
  \ and\ \bibinfo {author} {\bibfnamefont {D.}~\bibnamefont {Klemm}},\ }\href
  {\doibase 10.1088/0264-9381/17/2/310} {\bibfield  {journal} {\bibinfo
  {journal} {Class. Quant. Grav.}\ }\textbf {\bibinfo {volume} {17}},\ \bibinfo
  {pages} {399} (\bibinfo {year} {2000})},\ \Eprint
  {http://arxiv.org/abs/hep-th/9908022} {arXiv:hep-th/9908022} \BibitemShut
  {NoStop}%
\bibitem [{\citenamefont {Chamblin}\ \emph
  {et~al.}(1999{\natexlab{a}})\citenamefont {Chamblin}, \citenamefont
  {Emparan}, \citenamefont {Johnson},\ and\ \citenamefont
  {Myers}}]{Chamblin:1999tk}%
  \BibitemOpen
  \bibfield  {author} {\bibinfo {author} {\bibfnamefont {A.}~\bibnamefont
  {Chamblin}}, \bibinfo {author} {\bibfnamefont {R.}~\bibnamefont {Emparan}},
  \bibinfo {author} {\bibfnamefont {C.~V.}\ \bibnamefont {Johnson}}, \ and\
  \bibinfo {author} {\bibfnamefont {R.~C.}\ \bibnamefont {Myers}},\ }\href
  {\doibase 10.1103/PhysRevD.60.064018} {\bibfield  {journal} {\bibinfo
  {journal} {Phys.Rev.D}\ }\textbf {\bibinfo {volume} {60}},\ \bibinfo {pages}
  {064018} (\bibinfo {year} {1999}{\natexlab{a}})},\ \Eprint
  {http://arxiv.org/abs/hep-th/9902170} {arXiv:hep-th/9902170} \BibitemShut
  {NoStop}%
\bibitem [{\citenamefont {Chamblin}\ \emph
  {et~al.}(1999{\natexlab{b}})\citenamefont {Chamblin}, \citenamefont
  {Emparan}, \citenamefont {Johnson},\ and\ \citenamefont
  {Myers}}]{Chamblin:1999hg}%
  \BibitemOpen
  \bibfield  {author} {\bibinfo {author} {\bibfnamefont {A.}~\bibnamefont
  {Chamblin}}, \bibinfo {author} {\bibfnamefont {R.}~\bibnamefont {Emparan}},
  \bibinfo {author} {\bibfnamefont {C.~V.}\ \bibnamefont {Johnson}}, \ and\
  \bibinfo {author} {\bibfnamefont {R.~C.}\ \bibnamefont {Myers}},\ }\href
  {\doibase 10.1103/PhysRevD.60.104026} {\bibfield  {journal} {\bibinfo
  {journal} {Phys. Rev. D}\ }\textbf {\bibinfo {volume} {60}},\ \bibinfo
  {pages} {104026} (\bibinfo {year} {1999}{\natexlab{b}})},\ \Eprint
  {http://arxiv.org/abs/hep-th/9904197} {arXiv:hep-th/9904197} \BibitemShut
  {NoStop}%
\bibitem [{\citenamefont {Kubiznak}\ and\ \citenamefont
  {Mann}(2012)}]{Kubiznak:2012wp}%
  \BibitemOpen
  \bibfield  {author} {\bibinfo {author} {\bibfnamefont {D.}~\bibnamefont
  {Kubiznak}}\ and\ \bibinfo {author} {\bibfnamefont {R.~B.}\ \bibnamefont
  {Mann}},\ }\href {\doibase 10.1007/JHEP07(2012)033} {\bibfield  {journal}
  {\bibinfo  {journal} {JHEP}\ }\textbf {\bibinfo {volume} {07}},\ \bibinfo
  {pages} {033} (\bibinfo {year} {2012})},\ \Eprint
  {http://arxiv.org/abs/1205.0559} {arXiv:1205.0559 [hep-th]} \BibitemShut
  {NoStop}%
\bibitem [{\citenamefont {Gunasekaran}\ \emph {et~al.}(2012)\citenamefont
  {Gunasekaran}, \citenamefont {Mann},\ and\ \citenamefont
  {Kubiznak}}]{Gunasekaran:2012dq}%
  \BibitemOpen
  \bibfield  {author} {\bibinfo {author} {\bibfnamefont {S.}~\bibnamefont
  {Gunasekaran}}, \bibinfo {author} {\bibfnamefont {R.~B.}\ \bibnamefont
  {Mann}}, \ and\ \bibinfo {author} {\bibfnamefont {D.}~\bibnamefont
  {Kubiznak}},\ }\href {\doibase 10.1007/JHEP11(2012)110} {\bibfield  {journal}
  {\bibinfo  {journal} {JHEP}\ }\textbf {\bibinfo {volume} {11}},\ \bibinfo
  {pages} {110} (\bibinfo {year} {2012})},\ \Eprint
  {http://arxiv.org/abs/1208.6251} {arXiv:1208.6251 [hep-th]} \BibitemShut
  {NoStop}%
\bibitem [{\citenamefont {Altamirano}\ \emph {et~al.}(2013)\citenamefont
  {Altamirano}, \citenamefont {Kubiznak},\ and\ \citenamefont
  {Mann}}]{Altamirano:2013ane}%
  \BibitemOpen
  \bibfield  {author} {\bibinfo {author} {\bibfnamefont {N.}~\bibnamefont
  {Altamirano}}, \bibinfo {author} {\bibfnamefont {D.}~\bibnamefont
  {Kubiznak}}, \ and\ \bibinfo {author} {\bibfnamefont {R.~B.}\ \bibnamefont
  {Mann}},\ }\href {\doibase 10.1103/PhysRevD.88.101502} {\bibfield  {journal}
  {\bibinfo  {journal} {Phys. Rev. D}\ }\textbf {\bibinfo {volume} {88}},\
  \bibinfo {pages} {101502} (\bibinfo {year} {2013})},\ \Eprint
  {http://arxiv.org/abs/1306.5756} {arXiv:1306.5756 [hep-th]} \BibitemShut
  {NoStop}%
\bibitem [{\citenamefont {Altamirano}\ \emph {et~al.}(2014)\citenamefont
  {Altamirano}, \citenamefont {Kubiznak}, \citenamefont {Mann},\ and\
  \citenamefont {Sherkatghanad}}]{Altamirano:2014tva}%
  \BibitemOpen
  \bibfield  {author} {\bibinfo {author} {\bibfnamefont {N.}~\bibnamefont
  {Altamirano}}, \bibinfo {author} {\bibfnamefont {D.}~\bibnamefont
  {Kubiznak}}, \bibinfo {author} {\bibfnamefont {R.~B.}\ \bibnamefont {Mann}},
  \ and\ \bibinfo {author} {\bibfnamefont {Z.}~\bibnamefont {Sherkatghanad}},\
  }\href {\doibase 10.3390/galaxies2010089} {\bibfield  {journal} {\bibinfo
  {journal} {Galaxies}\ }\textbf {\bibinfo {volume} {2}},\ \bibinfo {pages}
  {89} (\bibinfo {year} {2014})},\ \Eprint {http://arxiv.org/abs/1401.2586}
  {arXiv:1401.2586 [hep-th]} \BibitemShut {NoStop}%
\bibitem [{\citenamefont {Banados}\ \emph {et~al.}(1993)\citenamefont
  {Banados}, \citenamefont {Henneaux}, \citenamefont {Teitelboim},\ and\
  \citenamefont {Zanelli}}]{Banados:1992gq}%
  \BibitemOpen
  \bibfield  {author} {\bibinfo {author} {\bibfnamefont {M.}~\bibnamefont
  {Banados}}, \bibinfo {author} {\bibfnamefont {M.}~\bibnamefont {Henneaux}},
  \bibinfo {author} {\bibfnamefont {C.}~\bibnamefont {Teitelboim}}, \ and\
  \bibinfo {author} {\bibfnamefont {J.}~\bibnamefont {Zanelli}},\ }\href
  {\doibase 10.1103/PhysRevD.48.1506, 10.1103/PhysRevD.88.069902} {\bibfield
  {journal} {\bibinfo  {journal} {Phys. Rev.}\ }\textbf {\bibinfo {volume}
  {D48}},\ \bibinfo {pages} {1506} (\bibinfo {year} {1993})},\ \bibinfo {note}
  {[Erratum: Phys. Rev.D88,069902(2013)]},\ \Eprint
  {http://arxiv.org/abs/gr-qc/9302012} {arXiv:gr-qc/9302012 [gr-qc]}
  \BibitemShut {NoStop}%
\bibitem [{\citenamefont {Banados}\ \emph {et~al.}(1992)\citenamefont
  {Banados}, \citenamefont {Teitelboim},\ and\ \citenamefont
  {Zanelli}}]{Banados:1992wn}%
  \BibitemOpen
  \bibfield  {author} {\bibinfo {author} {\bibfnamefont {M.}~\bibnamefont
  {Banados}}, \bibinfo {author} {\bibfnamefont {C.}~\bibnamefont {Teitelboim}},
  \ and\ \bibinfo {author} {\bibfnamefont {J.}~\bibnamefont {Zanelli}},\ }\href
  {\doibase 10.1103/PhysRevLett.69.1849} {\bibfield  {journal} {\bibinfo
  {journal} {Phys. Rev. Lett.}\ }\textbf {\bibinfo {volume} {69}},\ \bibinfo
  {pages} {1849} (\bibinfo {year} {1992})},\ \Eprint
  {http://arxiv.org/abs/hep-th/9204099} {arXiv:hep-th/9204099 [hep-th]}
  \BibitemShut {NoStop}%
\bibitem [{\citenamefont {Frassino}\ \emph {et~al.}(2015)\citenamefont
  {Frassino}, \citenamefont {Mann},\ and\ \citenamefont
  {Mureika}}]{Frassino:2015oca}%
  \BibitemOpen
  \bibfield  {author} {\bibinfo {author} {\bibfnamefont {A.~M.}\ \bibnamefont
  {Frassino}}, \bibinfo {author} {\bibfnamefont {R.~B.}\ \bibnamefont {Mann}},
  \ and\ \bibinfo {author} {\bibfnamefont {J.~R.}\ \bibnamefont {Mureika}},\
  }\href {\doibase 10.1103/PhysRevD.92.124069} {\bibfield  {journal} {\bibinfo
  {journal} {Phys. Rev.}\ }\textbf {\bibinfo {volume} {D92}},\ \bibinfo {pages}
  {124069} (\bibinfo {year} {2015})},\ \Eprint
  {http://arxiv.org/abs/1509.05481} {arXiv:1509.05481 [gr-qc]} \BibitemShut
  {NoStop}%
\bibitem [{\citenamefont {Kubiznak}\ \emph {et~al.}(2017)\citenamefont
  {Kubiznak}, \citenamefont {Mann},\ and\ \citenamefont
  {Teo}}]{Kubiznak:2016qmn}%
  \BibitemOpen
  \bibfield  {author} {\bibinfo {author} {\bibfnamefont {D.}~\bibnamefont
  {Kubiznak}}, \bibinfo {author} {\bibfnamefont {R.~B.}\ \bibnamefont {Mann}},
  \ and\ \bibinfo {author} {\bibfnamefont {M.}~\bibnamefont {Teo}},\ }\href
  {\doibase 10.1088/1361-6382/aa5c69} {\bibfield  {journal} {\bibinfo
  {journal} {Class. Quant. Grav.}\ }\textbf {\bibinfo {volume} {34}},\ \bibinfo
  {pages} {063001} (\bibinfo {year} {2017})},\ \Eprint
  {http://arxiv.org/abs/1608.06147} {arXiv:1608.06147 [hep-th]} \BibitemShut
  {NoStop}%
\bibitem [{\citenamefont {Mann}(1993)}]{PhysRevD.47.4438}%
  \BibitemOpen
  \bibfield  {author} {\bibinfo {author} {\bibfnamefont {R.~B.}\ \bibnamefont
  {Mann}},\ }\href {\doibase 10.1103/PhysRevD.47.4438} {\bibfield  {journal}
  {\bibinfo  {journal} {Phys. Rev. D}\ }\textbf {\bibinfo {volume} {47}},\
  \bibinfo {pages} {4438} (\bibinfo {year} {1993})}\BibitemShut {NoStop}%
\bibitem [{\citenamefont {Louis-Martinez}\ and\ \citenamefont
  {Kunstatter}(1995)}]{PhysRevD.52.3494}%
  \BibitemOpen
  \bibfield  {author} {\bibinfo {author} {\bibfnamefont {D.}~\bibnamefont
  {Louis-Martinez}}\ and\ \bibinfo {author} {\bibfnamefont {G.}~\bibnamefont
  {Kunstatter}},\ }\href {\doibase 10.1103/PhysRevD.52.3494} {\bibfield
  {journal} {\bibinfo  {journal} {Phys. Rev. D}\ }\textbf {\bibinfo {volume}
  {52}},\ \bibinfo {pages} {3494} (\bibinfo {year} {1995})}\BibitemShut
  {NoStop}%
\bibitem [{\citenamefont {Wald}(1993)}]{Wald:1993nt}%
  \BibitemOpen
  \bibfield  {author} {\bibinfo {author} {\bibfnamefont {R.~M.}\ \bibnamefont
  {Wald}},\ }\href {\doibase 10.1103/PhysRevD.48.R3427} {\bibfield  {journal}
  {\bibinfo  {journal} {Phys. Rev. D}\ }\textbf {\bibinfo {volume} {48}},\
  \bibinfo {pages} {3427} (\bibinfo {year} {1993})},\ \Eprint
  {http://arxiv.org/abs/gr-qc/9307038} {arXiv:gr-qc/9307038} \BibitemShut
  {NoStop}%
\bibitem [{\citenamefont {Myers}(1994)}]{Myers:1994sg}%
  \BibitemOpen
  \bibfield  {author} {\bibinfo {author} {\bibfnamefont {R.~C.}\ \bibnamefont
  {Myers}},\ }\href {\doibase 10.1103/PhysRevD.50.6412} {\bibfield  {journal}
  {\bibinfo  {journal} {Phys. Rev. D}\ }\textbf {\bibinfo {volume} {50}},\
  \bibinfo {pages} {6412} (\bibinfo {year} {1994})},\ \Eprint
  {http://arxiv.org/abs/hep-th/9405162} {arXiv:hep-th/9405162} \BibitemShut
  {NoStop}%
\bibitem [{\citenamefont {Cisterna}\ \emph {et~al.}(2020)\citenamefont
  {Cisterna}, \citenamefont {Henr\'\i{}quez-B\'aez},\ and\ \citenamefont
  {Oliva}}]{Cisterna:2019scr}%
  \BibitemOpen
  \bibfield  {author} {\bibinfo {author} {\bibfnamefont {A.}~\bibnamefont
  {Cisterna}}, \bibinfo {author} {\bibfnamefont {C.}~\bibnamefont
  {Henr\'\i{}quez-B\'aez}}, \ and\ \bibinfo {author} {\bibfnamefont
  {J.}~\bibnamefont {Oliva}},\ }\href {\doibase 10.1007/JHEP01(2020)052}
  {\bibfield  {journal} {\bibinfo  {journal} {JHEP}\ }\textbf {\bibinfo
  {volume} {01}},\ \bibinfo {pages} {052} (\bibinfo {year} {2020})},\ \Eprint
  {http://arxiv.org/abs/1909.05404} {arXiv:1909.05404 [hep-th]} \BibitemShut
  {NoStop}%
\bibitem [{\citenamefont {Chan}\ \emph {et~al.}(1996)\citenamefont {Chan},
  \citenamefont {Creighton},\ and\ \citenamefont {Mann}}]{Chan:1996sx}%
  \BibitemOpen
  \bibfield  {author} {\bibinfo {author} {\bibfnamefont {K.}~\bibnamefont
  {Chan}}, \bibinfo {author} {\bibfnamefont {J.}~\bibnamefont {Creighton}}, \
  and\ \bibinfo {author} {\bibfnamefont {R.~B.}\ \bibnamefont {Mann}},\ }\href
  {\doibase 10.1103/PhysRevD.54.3892} {\bibfield  {journal} {\bibinfo
  {journal} {Phys. Rev. D}\ }\textbf {\bibinfo {volume} {54}},\ \bibinfo
  {pages} {3892} (\bibinfo {year} {1996})},\ \Eprint
  {http://arxiv.org/abs/gr-qc/9604055} {arXiv:gr-qc/9604055} \BibitemShut
  {NoStop}%
\bibitem [{\citenamefont {Hawking}\ and\ \citenamefont
  {Horowitz}(1996)}]{Hawking:1995fd}%
  \BibitemOpen
  \bibfield  {author} {\bibinfo {author} {\bibfnamefont {S.}~\bibnamefont
  {Hawking}}\ and\ \bibinfo {author} {\bibfnamefont {G.~T.}\ \bibnamefont
  {Horowitz}},\ }\href {\doibase 10.1088/0264-9381/13/6/017} {\bibfield
  {journal} {\bibinfo  {journal} {Class. Quant. Grav.}\ }\textbf {\bibinfo
  {volume} {13}},\ \bibinfo {pages} {1487} (\bibinfo {year} {1996})},\ \Eprint
  {http://arxiv.org/abs/gr-qc/9501014} {arXiv:gr-qc/9501014} \BibitemShut
  {NoStop}%
\bibitem [{\citenamefont {Emparan}\ \emph {et~al.}(1999)\citenamefont
  {Emparan}, \citenamefont {Johnson},\ and\ \citenamefont
  {Myers}}]{Emparan:1999pm}%
  \BibitemOpen
  \bibfield  {author} {\bibinfo {author} {\bibfnamefont {R.}~\bibnamefont
  {Emparan}}, \bibinfo {author} {\bibfnamefont {C.~V.}\ \bibnamefont
  {Johnson}}, \ and\ \bibinfo {author} {\bibfnamefont {R.~C.}\ \bibnamefont
  {Myers}},\ }\href {\doibase 10.1103/PhysRevD.60.104001} {\bibfield  {journal}
  {\bibinfo  {journal} {Phys. Rev. D}\ }\textbf {\bibinfo {volume} {60}},\
  \bibinfo {pages} {104001} (\bibinfo {year} {1999})},\ \Eprint
  {http://arxiv.org/abs/hep-th/9903238} {arXiv:hep-th/9903238} \BibitemShut
  {NoStop}%
\end{thebibliography}%
\end{document}